\documentclass[preprint,12pt]{elsarticle}

\usepackage{amssymb}
\usepackage{amsthm}
\usepackage{amsmath}
\usepackage{bm}
\usepackage{bbm}
\usepackage{hyperref}

\journal{Physics Letters B}

\begin{document}

\begin{frontmatter}

\title{Perturbative static quark potential\\
in Maximal Abelian gauge}

\author{Matthias Berwein}
\ead{matthias.berwein@riken.jp}
\address{Nishina Center, RIKEN, Wako, Saitama 351-0198, Japan}
\author{Yukinari Sumino}
\ead{yukinari.sumino.a4@tohoku.ac.jp}
\address{Department of Physics, Tohoku University, 6-3, aza-Aoba, Aramaki, Aoba-ku, Sendai 980-8578, Japan}

\begin{abstract}
We calculate the static quark potential for an SU($N$) gauge theory in the Maximal Abelian gauge as well as its Abelian projection up to two loops in perturbation theory. We discuss its renormalization properties. The result is compared with a recent lattice result at $r \lesssim 0.5$~fm.

\end{abstract}

\begin{keyword}

QCD \sep static quark potential \sep Maximal Abelian gauge \sep perturbation theory \sep renormalization

\end{keyword}

\end{frontmatter}

\section{Introduction}

One of the most interesting features of Quantum Chromodynamics (QCD) is its peculiar behavior at low energies, where it displays confinement. Among the many explanations proposed for this phenomenon, one of the more interesting approaches utilizes its similarity to the behavior of a magnetic field in a type II superconductor~\cite{tHooft:1981bkw}. The magnetic field cannot penetrate the superconductor except for narrow flux tubes, much like the chromoelectric field extends only between confined particles, and if there existed elementary magnetic charges, they would be confined inside a superconductor much like quarks in the vacuum. But whereas a superconductor forms through the condensation of electrons, there are no corresponding elementary chromomagnetic charges in QCD. Instead, it is argued that these are dynamically generated by the gluon field, and that they exist in a condensed state in the QCD vacuum.

This idea has already been extensively studied on the lattice using a particular gauge fixing, the so-called Maximal Abelian (MA) gauge~\cite{Kronfeld:1987ri}, in which chromomagnetic monopoles are topologically generated. Evidence has been found on the lattice for both the existence~\cite{Kronfeld:1987vd} of these monopoles and their condensation~\cite{Kronfeld:1987ri} below a critical temperature. A particularly clear picture emerges from the study of the static quark potential. The MA gauge treats gluons belonging to the maximal Abelian subgroup of SU$(N)$ (i.e., gluons with diagonal generators) differently from the other gluons, and the monopoles are all contained in this diagonal part of the gluon field. Separating the diagonal from the off-diagonal contributions to the potential, it appears that the linear behavior of the potential at long distances, which is responsible for confinement, is displayed solely by the diagonal part (in particular by the monopole contribution), while the off-diagonal part plays a minor role~\cite{Stack:1994wm} (or~\cite{Sakumichi:2014xpa} for a more recent study). In fact, neglecting the off-diagonal contribution completely does not change the qualitative behavior of the potential, a feature that is called Abelian dominance. While it is possible that these phenomena may just be coincidental artifacts of the gauge fixing, a common view is that the MA gauge is particularly convenient in organizing universal features of QCD that might be obscured in other gauges.

There is an interesting detail in this result. The linear behavior of the monopole contribution seems to extend even to short distances, where the overall behavior of the potential is Coulombic. In this region, the potential may be studied in perturbation theory. Currently, the static potential is computed up to three loops in perturbative QCD in the Feynman gauge~\cite{Anzai:2009tm} and in the general covariant gauge~\cite{Smirnov:2009fh}. To our knowledge, however, there exists no perturbative computation of the potential in the MA gauge. In this paper, in order to expand the picture, we conduct a two-loop calculation of the potential as well as its Abelian projection in the MA gauge. 

In section~\ref{MAG} we briefly discuss the particularities of MA gauge fixing, section~\ref{SQP} then reviews the necessary information on the static quark potential, while our results are presented in section~\ref{Res}.

\section{The Maximal Abelian gauge}
\label{MAG}

The Maximal Abelian gauge in the continuum is defined as the gauge that minimizes the following functional over the off-diagonal gluonic degrees of freedom
\begin{equation}
 \int d^4x\,A_\mu^a(x)A^{a\,\mu}(x)\,.
 \label{functional}
\end{equation}
We use here the notation that the letters $a$, $b$, $c$, $\dots$ denote off-diagonal degrees of freedom, while $i$, $j$, $k$, $\dots$ stand for diagonal ones. In case both are meant, capitals $A$, $B$, $C$, $\dots$ will be used. For instance, in the case of SU(3) $A\in\{1,\dots,8\}$, and with the standard Gell-Mann parametrization $a\in\{1,2,4,5,6,7\}$ and $i\in\{3,8\}$.

Performing an infinitesimal gauge transformation with parameter $g\theta^A$,
\begin{equation}
 A_\mu^A(x)\to A_\mu^A(x)+D_\mu^{AB}\theta^B(x)+\mathcal{O}(\theta^2)\,,
\end{equation}
where $D_\mu^{AB}$ is the covariant derivative in the adjoint representation, one obtains the local gauge condition
\begin{equation}
 d_\mu^{ab}A^{b\,\mu}(x)\equiv\left(\partial_\mu\delta^{ab}-gf^{abi}A_\mu^i(x)\right)A^{b\,\mu}(x)=0\,.
 \label{MAgauge-cond}
\end{equation}
In fact, this is only an extremal condition, not necessarily minimal, and it should also be noted that the functional~\eqref{functional} is not positive definite in Minkowski spacetime. However, the perturbative solution to Eq.~\eqref{MAgauge-cond} after a Wick rotation to Euclidean spacetime corresponds to a minimal condition, since the dominating second order term of the gauge transformed functional in the limit $g\to0$ is given by $|\partial_\mu\theta^a|^2\ge0$. Since the lattice calculations we want to compare to are performed in Euclidean spacetime, the Minkowskian analog~\eqref{MAgauge-cond} gives the correct gauge condition.

This condition fixes only $N(N-1)$ of the $N^2-1$ degrees of freedom, corresponding exactly to the number of off-diagonal gluons. The reason for this is that the above functional, as well as the local gauge condition itself, are invariant under purely diagonal gauge transformations $\theta^a=0$, which form a $\mathrm{U}(1)^{N-1}$ subgroup of SU($N$) (i.e., the maximal Abelian subgroup). In order to fix this residual gauge freedom, we choose the standard condition of covariant gauges for the diagonal gluons:
\begin{equation}
 \partial_\mu A^{i\,\mu}(x)=0\,,
 \label{MAgauge-sec-cond}
\end{equation}
although any other condition would be equally valid. In fact, with this choice it is possible to combine both conditions in one expression:
\begin{equation}
 d_\mu^{AB}A^{B\,\mu}(x)=0\,,
\end{equation}
where $d$, as defined above, is the covariant derivative of the residual gauge group.

We choose the Faddeev-Popov method to fix the gauge in the path integral. The reason for this is that then the gauge fixing parameter $\xi$ has a clear interpretation in terms of the delta functions used to enforce the gauge conditions:
\begin{equation}
 \delta(x)=\lim_{\xi\to0}\frac{1}{\sqrt{2\pi i\xi}}\exp\left[\frac{i}{2\xi}x^2\right]\,.
\end{equation}
Concerns have been raised~\cite{Min:1985bx,Kondo:2000ey} regarding the renormalizability of fixing the MA gauge in this way. Accordingly, most publications choose BRST gauge fixing, which ensures renormalizability, at the cost of obscuring the meaning of $\xi$. Furthermore, in these generalized MA gauges an additional parameter is introduced for interaction terms involving four ghost fields, which means that also the ghost part of the Lagrangian looses its interpretation as a functional determinant. In this letter, however, we are not interested in generalizations of the MA gauge, as the particular features involving monopoles are expected to appear only if the gauge condition is exactly satisfied. The problematic diagrams, whose divergences the additional terms in BRST gauge fixing are supposed to cancel, do not enter our calculation at the two-loop level, and it is not clear whether they contribute at all at higher orders. However, other finite contributions to the two-loop result are affected by the choice of gauge fixing, which is why it is important to point out here that we are using Faddeev-Popov instead of BRST. In order to distinguish the off-diagonal~\eqref{MAgauge-cond} from the diagonal conditions~\eqref{MAgauge-sec-cond}, we use separate parameters $\xi$ and $\eta$, respectively.

The resulting gauge fixing Lagrangian is thus given by
\begin{align}
 \mathcal{L}_{gf}={}&\frac{1}{2\xi}\left(d_\mu^{ab}A^{b\,\mu}\right)\left(d_\nu^{ac}A^{c\,\nu}\right)+\frac{1}{2\eta}\left(\partial_\mu A^{i\,\mu}\right)\left(\partial_\nu A^{i\,\nu}\right)\notag\\
 &-\bar{c}^Ad_\mu^{AB}D^{BC\,\mu}c^C+\bar{c}^agf^{abi}A_\mu^bD^{iC\,\mu}c^C\,.
\end{align}
Due to the quadratic nature of the gauge condition, the new terms in the gauge fixing Lagrangian affect not only the gluon propagators but also the vertices with two off-diagonal gluons and one or two diagonal gluons. Apart from those extra terms, the Feynman rules for gluons are the same as in covariant gauge. The ghost sector shows more differences to covariant gauge, in particular it also contains vertices with two ghost and two gluon fields.

The new terms in the three-gluon and four-gluon vertices have a $1/\xi$ coefficient that formally diverges in the $\xi\to0$ limit. However, it can be shown that these terms always cancel when all the diagrams contributing at a certain loop order are included. While it would be possible to formulate Feynman rules that take these cancellations explicitly into account and contain no $1/\xi$ terms, it is simpler to work with the standard rules, and the cancellation of $1/\xi$ terms can be used as a cross check.

\section{The static quark potential}
\label{SQP}

The static quark potential\footnote{Strictly speaking, the quantity defined in this way is called the static energy, while the static quark potential as defined in the effective theory of potential non-relativistic QCD~\cite{Pineda:1997bj,Brambilla:1999xf} differs from it by ultrasoft corrections that start to contribute at three loop order.} is given by the expectation value of a rectangular Wilson loop of spatial extent $\bm{r}$ and temporal extent $T$ (not to be confused with the color matrices $T^A$), according to the relation
\begin{equation}
 V_{\rm QCD}(r)=\lim_{T\to\infty}\frac{i}{T}\ln\frac{1}{N}\mathrm{Tr}\left\langle \mathcal{P}\exp\left[ig\int_\square dx^\mu A_\mu^A(x)T^A\right]\right\rangle\,,
\end{equation}
where $\mathcal{P}$ denotes path ordering, and the large time limit is understood after a Wick rotation $T=-iT_E$, $T_E\to\infty$. The Abelian projection is defined accordingly by including only diagonal gluons in the path ordered exponential.

The large time limit can be carried out before calculating any diagram by dropping the spatial parts of the Wilson line contour and assigning a propagator $i/(k_0+i\epsilon)$ to each part of the contour that appears in loops. This propagator has its origin in the momentum space representation of the theta functions facilitating the path ordering of gauge fields along the Wilson line. It can also be interpreted as a static quark propagator. A momentum is assigned to each section of the contour as well, which is conserved at every point where a gluon connects to the contour. Apart from loop momenta, there is one external momentum $(0,\bm{q})$ exchanged between the quark and the antiquark line, and the potential in position space is given as the $(D-1)$-dimensional Fourier transform of the momentum space result with respect to this momentum.

Taking the logarithm of a Wilson line average can be carried out at the level of Feynman diagrams. According to the exponentiation theorem for Wilson line operators~\cite{Dotsenko:1979wb,Gatheral:1983cz}, the logarithm is given by a subset of the original diagrams with modified coefficients. For that purpose, each diagram is written as the product of a color part, which contains the normalized trace over color matrices ordered along the contour and structure constants from vertices, and a kinematic part, which includes everything else, in particular the loop integrals over propagators and vertex functions. The modifications are carried out solely in the color part as a set of subtractions according to rules determined by the geometry of the diagram. For the derivation of this theorem using the replica trick~\cite{Gardi:2010rn}, it does not matter if diagonal and off-diagonal indices are distinguished or not. In general, the subset of diagrams with non-vanishing coefficients in the logarithm is considerably smaller than the set of original diagrams.

For the calculation of the color factors, the following Fierz identities are instrumental:
\begin{align}
 T_{IJ}^aT_{KL}^a&=\frac{1}{2}\left(\delta_{IL}\delta_{KJ}-\delta_{IJKL}\right)\,,\\
 T_{IJ}^iT_{KL}^i&=\frac{1}{2}\left(\delta_{IJKL}-\frac{1}{N}\delta_{IJ}\delta_{KL}\right)\,,
\end{align}
where we used capitals $I$, $J$, $\dots$, to distinguish matrix indices from color indices, and the generalized Kronecker symbol with more than two indices is defined to be 1 only if all the indices are identical and 0 otherwise. The sum of both expressions reproduces the known Fierz identity for the full set of generators, which can be derived from the orthogonality relation $\mathrm{Tr}[T^AT^B]=\delta^{AB}/2$ and the fact that the generators together with the unit matrix form a basis for the space of $N\times N$ matrices. In analogous fashion, the subset of diagonal generators (including the unit matrix) forms a basis for the space of diagonal $N\times N$ matrices, from which follows the second identity.

The calculation of the color coefficients then reduces to contracting indices of products of (generalized) Kronecker deltas, which shows that each coefficient is given as a function of $N$. There also exist diagonal and off-diagonal analogs of the quadratic Casimir:
\begin{align}
 T^aT^a&=\frac{N-1}{2}\mathbbm{1}_N\,,&T^iT^i&=\frac{N-1}{2N}\mathbbm{1}_N\,.
 \label{Casimirs}
\end{align}
The off-diagonal Casimir is larger than the diagonal one by a factor of $N$, which explains qualitatively the suppression of the Abelian projection with respect to the full potential found in our result (see below) simply through Casimir scaling.

\section{Result}
\label{Res}

While the one-loop calculation involves only integrals that can be solved by hand with the usual Feynman parameter methods, at two-loop order this already becomes inadvisable. In particular, the different Feynman rules for diagonal and off-diagonal gluons and ghosts increase the number of diagrams by an order of magnitude compared to the calculation in Feynman gauge. Therefore, we have written a code using the popular Laporta algorithm~\cite{Laporta:2001dd} to reduce the integrals down to five master integrals via integration-by-parts relations. The master integrals themselves are already known from previous calculations of the static potential~\cite{Schroder:1999sg}, while the steps leading to the generation of the integrals had to be modified to account for the particularities of the MA gauge. The calculation of the color coefficients using the replica trick and the Fierz identities is also relatively straightforward to implement. Details of our computation will be described elsewhere. 

Our result for the full potential agrees with the known result (cf.~\cite{Fischler:1977yf,Billoire:1979ih} for the one-loop and~\cite{Schroder:1999sg} for the two-loop order). It is independent of both gauge fixing parameters $\xi$ and $\eta$ due to the gauge invariance of the Wilson loop. The Abelian projection is only invariant under the diagonal subgroup, thus we only expect it to be independent of $\eta$, which indeed is the case. Starting at one-loop order, the Abelian projection becomes explicitly dependent on $\xi$. In addition, at the two-loop level there appear UV divergences that are not removed by charge renormalization and require to also consider $\xi$ as a renormalized parameter in order to generate counterterms that eliminate these divergences.

Both the full potential and the Abelian projection can be written schematically in the form
\begin{align}
 V_{\rm QCD}(|\bm{q}|)={}&-\frac{4\pi C_F\alpha_s(|\bm{q}|)}{\bm{q}^2}\left[1+\frac{\alpha_s(|\bm{q}|)}{4\pi}\left(a_1+b_1\xi(|\bm{q}|)+c_1\xi(|\bm{q}|)^2\right)\right.\notag\\
 &+\left.\left(\frac{\alpha_s(|\bm{q}|)}{4\pi}\right)^2\left(a_2+b_2\xi(|\bm{q}|)+c_2\xi(|\bm{q}|)^2+d_2\xi(|\bm{q}|)^3\right)+\mathcal{O}\left(\alpha_s^3\right)\right]\,.
\label{Vmom}
\end{align}
The individual constants are listed below for the full potential
\begin{align}
 a_1^\mathrm{(full)}={}&\frac{31}{9}N-\frac{10}{9}n_f\,,\qquad b_1^\mathrm{(full)}=c_1^\mathrm{(full)}=b_2^\mathrm{(full)}=c_2^\mathrm{(full)}=d_2^\mathrm{(full)}=0\,,\\
 a_2^\mathrm{(full)}={}&\left(\frac{4343}{162}+4\pi^2-\frac{\pi^4}{4}+\frac{22}{3}\zeta(3)\right)N^2-\left(\frac{899}{81}+\frac{28}{3}\zeta(3)\right)Nn_f\notag\\
 &-\left(\frac{55}{6}-8\zeta(3)\right)\frac{N^2-1}{2N}n_f+\frac{100}{81}n_f^2\,,
\end{align}
and for the Abelian projection
\begin{align}
 a_1^\mathrm{(AP)}={}&\frac{205}{36}N-\frac{10}{9}n_f\,,\qquad b_1^\mathrm{(AP)}=\frac{3}{2}N\,,\qquad c_1^\mathrm{(AP)}=\frac{1}{4}N\,,\\
 a_2^\mathrm{(AP)}={}&\left(\frac{90391}{1296}-\frac{57}{8}\zeta(3)\right)N^2+\left(\frac{347}{24}-\frac{115}{4}\zeta(3)\right)N\notag\\
 &-\left(\frac{1736}{81}+4\zeta(3)\right)Nn_f-\left(\frac{55}{6}-8\zeta(3)\right)\frac{N^2-1}{2N}n_f+\frac{100}{81}n_f^2\,,\\
 b_2^\mathrm{(AP)}={}&\left(\frac{211}{16}+\frac{5}{4}\zeta(3)\right)N^2+\left(\frac{367}{24}-\frac{7}{2}\zeta(3)\right)N-\frac{5}{3}Nn_f\,,\\
 c_2^\mathrm{(AP)}={}&\left(\frac{191}{48}-\frac{1}{8}\zeta(3)\right)N^2+\left(\frac{145}{24}+\frac{1}{4}\zeta(3)\right)N\,,\\
 d_2^\mathrm{(AP)}={}&\frac{9}{16}N^2+\frac{7}{8}N\,,
\end{align}
where $n_f$ is the number of massless quarks. The quadratic Casimir of the full group is $C_F^\mathrm{(full)}=(N^2-1)/(2N)$, while for the Abelian projection the result from Eq.~\eqref{Casimirs} has to be inserted, $C_F^\mathrm{(AP)} = (N-1)/(2N)$.

The strong coupling constant $\alpha_s (\mu)$ is defined in the $\overline{\rm MS}$ scheme, and its running can be found in the literature. The scale dependence of the gauge fixing parameter $\xi$ is given by
\begin{equation}
 \xi(|\bm{q}|)=\xi(\mu)\left[1-\frac{\alpha_s(\mu)}{4\pi}\left(\frac{\zeta_0^{(-1)}}{\xi(\mu)}+\zeta_0^{(0)}+\zeta_0^{(1)}\xi(\mu)\right)\ln\frac{\bm{q}^2}{\mu^2}+\mathcal{O}\left(\alpha_s^2\right)\right]\,,
\end{equation}
with
\begin{equation}
 \zeta_0^{(-1)}=3\,,\qquad \zeta_0^{(0)}=-\frac{13}{6}N+3+\frac{2}{3}n_f\,,\qquad \zeta_0^{(1)}=\frac{1}{2}N+1\,.
\end{equation}
We have checked these coefficients in an independent calculation of the off-diagonal gluon self-energy and found consistency. They also agree with a calculation of the associated anomalous dimension found in~\cite{Gracey:2005vu}, up to terms in $\zeta_0^{(1)}$ that can be traced back to the different gauge fixing method.

\begin{figure}[t]
 \begin{center}
 \includegraphics[width=9cm]{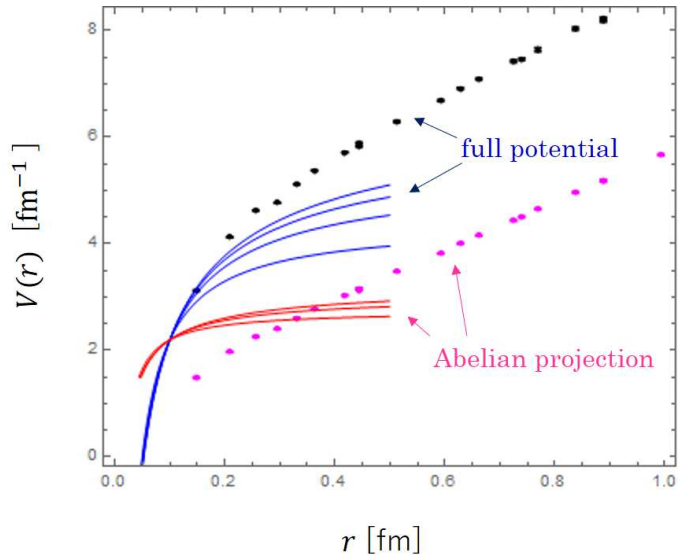}
 \end{center}
 \caption{Comparison of the lattice computation (data points with error bars) and the perturbative computation (solid lines) of the static potential. An arbitrary $r$-independent constant is added to each potential. The blue solid lines represent, from bottom to top, the tree, 1-loop, 2-loop and 3-loop full potentials. The red solid lines represent, from bottom to top, the tree, 1-loop and 2-loop Abelian-projected potentials in the MA gauge.}
 \label{fig1}
\end{figure}

The momentum-space potential~\eqref{Vmom} is Fourier transformed to the position-space potential $V_{\rm QCD}(r)$ after expanding $V_{\rm QCD}(|\bm{q}|)$ in $\alpha_s(\mu)$. In Fig.~\ref{fig1} we compare our result of the Abelian projection of the perturbative potential in the MA gauge with the corresponding one from a recent lattice computation~\cite{Sakumichi:2015rfa}, for SU(3) and in the quenched approximation ($n_f=0$). For reference, the results of the full potentials are also shown. The perturbative result at three-loop order and lattice data show a reasonable agreement up to around $r  \lesssim 0.2$--$0.25$~fm for the full potential.\footnote{The agreement is better in the case that the continuum limit of the lattice data is taken~\cite{Anzai:2009tm}. Namely, the shortest-distance data points of the lattice potentials in Fig.~\ref{fig1} may have substantial corrections from discretization effects.} In plotting the perturbative potentials, we use $r_0\, \Lambda_{\overline{\rm MS}}=0.574$~\cite{Sumino:2005cq}, where $r_0=0.5$~fm denotes the Sommer scale, and set $ \Lambda_{\overline{\rm MS}}/\mu=0.07$ and $\alpha_s(\mu)=0.165$. We set the renormalized $\xi=0$ in the Abelian projection of the potential.

We find no qualitative difference between the full potential and its Abelian projection. Both are Coulombic in shape with logarithmic corrections, and the coefficients $a_1$ and $a_2$ are of the same order of magnitude. Hence, the main difference comes from an overall factor $N+1$ that is determined by the ratio of the respective quadratic Casimirs. While it should be noted that the phenomenon of Abelian dominance is only observed for long distances, which lie outside the reach of perturbation theory, and a difference in the behavior of the full potential and its Abelian projection at short distances has also been pointed out in~\cite{Sakumichi:2014xpa}, there still is no trace of the monopole contribution in our perturbative result. This may be related to the following issue.

The fact that $\zeta_0^{(-1)}\neq0$ poses a serious problem, as it makes it impossible to set $\xi(\mu)=0$ for all the scales $\mu$. In particular, we cannot set the bare $\xi$ to zero while keeping the Abelian-projected potential finite. Recall that the MA gauge condition is only satisfied if the bare $\xi$ exactly vanishes and that the particular features of this gauge such as monopole formation are not necessarily expected in the modified gauge for $\xi\neq0$. An alternative would be to leave $\xi$ unrenormalized and set the bare $\xi=0$, at the cost of keeping uncanceled UV divergences in the result for the Abelian projection. Whether these can be taken care of in some other fashion, e.g., by explicitly including monopole contributions, remains a very interesting but as yet unsolved problem. These issues may also explain the discrepancy we find in our result compared to lattice data of the Abelian projection at short distances, where the full potential already shows rather good agreement.

\section*{Acknowledgements}

The authors are grateful to N.~Sakumichi and H.~Suganuma for providing their lattice data. The authors also thank K.~Kondo, S.~Sasaki and H.~Suganuma for fruitful discussion. This work was supported in part by Grant-in-Aid for JSPS Research Fellow (No.~16F16797). M.B.\ is supported by the Special Postdoctoral Researcher program of RIKEN.

\bibliographystyle{elsarticle-num} 
\bibliography{Ref}

\end{document}